\documentclass[aps,prd,nofootinbib,11pt]{revtex4}
\usepackage{amsmath}
\usepackage{graphicx}
\usepackage{dcolumn}
\usepackage{bm}
\usepackage{amssymb}
\usepackage{latexsym}

\newcommand{\be}{\begin{equation}}
\newcommand{\ee}{\end{equation}}
\newcommand{\bq}{\begin{eqnarray}}
\newcommand{\eq}{\end{eqnarray}}

\begin{document}

\title{New agegraphic dark energy as a rolling tachyon}

\author{Jinglei Cui}
\affiliation{Department of Physics, Northeastern University,
Shenyang 110004, People's Republic of China}
\author{Li Zhang}
\affiliation{Department of Physics, Northeastern University,
Shenyang 110004, People's Republic of China}
\author{Jingfei Zhang}
\affiliation{Department of Physics, Northeastern University,
Shenyang 110004, People's Republic of China} 
\author{Xin Zhang}
\affiliation{Department of Physics, Northeastern University,
Shenyang 110004, People's Republic of China} 

\begin{abstract}

Combining the general relativity and the uncertainty relation in
quantum mechanics, the energy density of quantum fluctuations of
space-time can be viewed as dark energy. The so-called agegraphic
dark energy model is just based on this viewpoint, in which the age
of the universe is introduced as the length measure. Recently, the
new agegraphic dark energy model was proposed, where the dynamical
dark energy is measured by the conformal age of the universe. On the
other hand, scalar-field dark energy models like tachyon are often
regarded as an effective description of some underlying theory of
dark energy. In this paper, we show that the new agegraphic dark
energy can be described completely by a tachyon scalar-field. We
thus reconstruct the potential and the dynamics of the tachyon
scalar-field, according to the evolution of the new agegraphic dark
energy.
\end{abstract}

\pacs{98.80.-k, 95.36.+x}

\maketitle

Nowadays it is strongly believed that the universe is currently
undergoing an accelerated expansion. The astronomical observations,
such as searches for type Ia supernovae (SNIa)~\cite{Riess98},
observations of large scale structure (LSS)~\cite{Tegmark04} and
measurements of the cosmos microware background (CMB)
anisotropy~\cite{Spergel03}, have provided main evidence for this
cosmic acceleration. It is the most accepted idea that this cosmic
acceleration is caused by a mysterious energy component, usually
dubbed ``dark energy", with negative pressure, whose energy density
has been a dominative power of the universe.  A distinct feature of
dark energy is that it is unclustered, differing from baryons and
nonbaryonic cold dark matter that have feature of gravitational
clustering. The combined analysis of cosmological observations
indicates that the universe is spatially flat, and consists of about
70\% dark energy, 30\% dust matter (cold dark matter plus baryons),
and negligible radiation. Though the nature of dark energy is still
unknown, we can affirm that the ultimate fate of the universe is
determined by the nature of dark energy. Currently it is the central
problem to reveal the nature of dark energy in the research of both
cosmology and theoretical physics. To interpret or describe its
properties, many candidates for dark energy have been proposed. The
famous cosmological constant introduced first by Einstein in 1917 is
the simplest candidate for dark energy. However, the cosmological
constant scenario has to face the so-called ``fine-tuning problem''
and ``coincidence problem''~\cite{lamada}. The cosmological constant
problem is essentially an issue of quantum gravity, because of
concerning the vacuum expectation value of some quantum fields.
However we have no a complete theory of quantum gravity yet. But
theorists have been making lots of efforts to try to resolve the
cosmological constant problem by various means.

Another candidate for dark energy is dynamical scalar field model
whose energy density is associated with a slowly varying scalar
field. The scalar field models can be viewed as an effective
description of the underlying theory of dark energy. An example of
scalar-field dark energy is the so-called
``quintessence"~\cite{quintessence}, a scalar field slowly evolving
down its potential. Provided that the evolution of the field is slow
enough, the kinetic energy density is less than the potential energy
density, giving rise to the negative pressure responsible to the
cosmic acceleration. So far, besides quintessence, a wide variety of
scalar-field dark energy models have been studied including
$K$-essence~\cite{kessence}, phantom~\cite{phantom},
tachyon~\cite{tachyon}, quintom~\cite{quintom}, ghost
condensate~\cite{ghost}, etc.. In addition, there are other
proposals on dark energy such as interacting dark energy
models~\cite{intde}, brane world models~\cite{brane}, and Chaplygin
gas models~\cite{cg}, and so on. However, one should realize that
almost these models are settled at the phenomenological level.

In recent years, many string theorists have devoted to sheding light
on the cosmological constant or dark energy within the string
framework. The famous KKLT model is a typical example, which tries
to construct metastable de Sitter vacua in light of type IIB string
theory. Furthermore, string landscape idea has been proposed for
shedding light on the cosmological constant problem based upon the
anthropic principle and multiverse speculation. Besides, according
to some principles of quantum gravity, we still can make some
attempts to probe the properties of dark energy within the
fundamental theory framework although a complete theory of quantum
gravity is not available today. The holographic dark energy
model~\cite{Li04,holofit} and the agegraphic dark energy
model~\cite{Cai1,Cai2} are just appropriate examples, which are
originated from some considerations of the quantum theory of
gravity. The holographic dark energy model is constructed in the
light of the holographic principle of quantum gravity theory, and
the agegraphic dark energy model takes into account the well-known
Heisenberg uncertainty relation of quantum mechanics together with
the gravitational effect in general relativity. In this regard, the
holographic and agegraphic dark energy scenarios may possess some
significant features of an underlying theory of dark energy.

As mentioned above, the scalar field models can be regarded as the
effective models of the underlying theory. In this sense, it is
interesting and meaningful to study how the scalar field models can
be used to describe dark energy scenarios possessing some
significant features of the quantum gravity theory, such as
holographic and agegraphic dark energy models. In this direction,
many works have been done, see, e.g., \cite{holorec}. In this paper,
we intend to reconstruct the potential and dynamics of tachyon
scalar-field from the new agegraphic dark energy model.

Let us first briefly review the models of agegraphic dark energy.
Based on quantum field theory, an ultraviolet (UV) cutoff is related
to an infrared (IR) cutoff due to the limit set by formation of a
black hole. It means that if $\rho_{\Lambda}$ is the quantum
zero-point energy density caused by UV cutoff, the total energy in a
region of size $L$ should not exceed the mass of a black hole of the
same size. The largest IR cut-off $L$ is chosen by saturating the
inequality $L^3\rho_\Lambda\leqslant L M_p^2$, so that we get the
holographic dark energy density~\cite{Li04}
\begin{equation}
\rho_{\Lambda}=3c^2M_p^2L^{-2}~,\label{de}
\end{equation}
where $c$ is a numerical constant, and $M_p\equiv 1/\sqrt{8\pi G}$
is the reduced Planck mass. Now the problem we look out on is how to
choose the IR cutoff. If the current Hubble horizon acts as the IR
cutoff, the equation of state of dark energy is the same as that of
dark matter, and it cannot lead to a universe with accelerating
expansion. For similar reason, the particle horizon of the universe
also cannot be used. Although the even horizon of the universe
acting as the IR cutoff can reach good conclusion that the universe
can undergo an accelerated expansion, and the vacuum energy density
in this case can fit the data well, it is surprising for us that the
current dark energy density is determined by the future evolution of
the universe, rather than the past of the universe.

Following the line of quantum fluctuations of spacetime,
K$\acute{a}$rolyh$\acute{a}$zy~\cite{r1} proposed that the distance
$t$ in Minkowski spacetime cannot be known to a better accuracy than
\begin{equation}
\delta t = \lambda t ^{2/3} _{p} t ^{1/3}, \label{eq1}
\end{equation}
where $\lambda$ is a dimensionless constant of order unity. We use
the units $\hbar = c = k_{B} = 1$ throughout this work. Thus $l_{p}
= t_{p} = 1/M_{p}$ with $l_{p}$ and $t_{p}$ being the reduced Planck
length and time, respectively.

The K\'{a}rolyh\'{a}zy relation~(\ref{eq1}) together with the
time-energy uncertainty relation enables one to estimate a quantum
energy density of the metric fluctuations of Minkowski
spacetime~\cite{r2,r3}. Following~\cite{r2,r3}, with respect to
Eq.~(\ref{eq1}) a length scale $t$ can be known with a maximum
precision $\delta t$ determining thereby a minimal detectable cell
$\delta t^3\sim t_p^2 t$ over a spatial region $t^3$. Such a cell
represents a minimal detectable unit of spacetime over a given
length scale $t$. If the age of the Minkowski spacetime is $t$, then
over a spatial region with linear size $t$ (determining the maximal
observable patch) there exists a minimal cell $\delta t^3$ the
energy of which due to time-energy uncertainty relation cannot be
smaller than
\begin{equation}
 E_{\delta t^3}\sim t^{-1}~.\label{eq2}
 \end{equation}
Therefore, the energy density of metric fluctuations of
 Minkowski spacetime is given by~\cite{r3,r2}
 \begin{equation}
 \rho_q\sim\frac{E_{\delta t^3}}{\delta t^3}\sim
 \frac{1}{t_p^2 t^2}\sim\frac{M_p^2}{t^2}~.\label{eq3}
 \end{equation}

In the original version of agegraphic dark energy model~\cite{Cai1},
the time scale $t$ in Eq.~(\ref{eq3}) is chosen to be the age of the
universe
\begin{equation}
 t=\int_0^a\frac{da}{Ha},\label{eq4}
 \end{equation}
where $a$ is the scale factor of our universe; $H\equiv\dot{a}/a$
 is the Hubble parameter; a dot denotes the derivative with respect
 to cosmic time.

To avoid some internal inconsistencies in the original model, the
so-called ``new agegraphic dark energy'' (NADE) model~\cite{Cai2}
was proposed, where the time scale in Eq.~(\ref{eq3}) is chosen to
be the conformal time $\eta$ instead of the age of the universe,
\begin{equation}
 \eta\equiv\int\frac{dt}{a}=\int\frac{da}{a^2H}.\label{eq5}
 \end{equation}
Consequently the energy density of NADE is
\begin{equation}
 \rho_q=\frac{3n^2M_p^2}{\eta^2},\label{eq6}
 \end{equation}
where the numerical factor $3n^2$ is introduced to parameterize some
uncertainties, such as the species of quantum fields in the
universe, the effect of curved spacetime (since the energy density
is derived for Minkowski spacetime), and so on.

We consider a spatially flat Friedmann-Robertson-Walker (FRW)
universe with dark energy and pressureless matter, so the
corresponding Friedmann equation reads
\begin{equation}
H^2=\frac{1}{3M_p^2}\left(\rho_m+\rho_q\right).\label{eq7}
\end{equation}
For NADE, the corresponding fractional energy density is given by
 \begin{equation}
 \Omega_q=\frac{n^2}{H^2\eta^2}.\label{eq8}
 \end{equation}

By using  Eqs.~(\ref{eq5})$-$(\ref{eq7}) and the energy conservation
equation $\dot{\rho}_m+3H\rho_m=0$, we find that the equation of
motion for $\Omega_q$ is given by
\begin{equation}
 \frac{d\Omega_q}{da}=\frac{\Omega_q}{a}\left(1-\Omega_q\right)
 \left(3-\frac{2}{n}\frac{\sqrt{\Omega_q}}{a}\right).\label{eq9}
 \end{equation}
Also, we can rewrite Eq.~(\ref{eq9}) as
\begin{equation}
\frac{d\Omega_{q}}{dz} =
-\Omega_{q}(1-\Omega_{q})\left[3(1+z)^{-1}-\frac{2}{n}\sqrt{\Omega_{q}}\right],\label{eq10}
\end{equation}
where $z = 1/a - 1$ is the redshif of the universe.

From the energy conservation equation $\dot{\rho_{q}} + 3H(\rho_{q}
+ p_{q}) = 0$, as well as Eqs.~(\ref{eq6}) and (\ref{eq8}), it is
easy to find that the equation-of-state of NADE, $w_{q} \equiv
p_{q}/\rho_{q}$, is given by
\begin{equation}
w_q=-1+\frac{2}{3n}(1+z)\sqrt{\Omega_q}.~\label{eq11}
\end{equation}

\begin{center}
 \begin{figure}[htbp]
 \centering
 \includegraphics[width=0.6\textwidth]{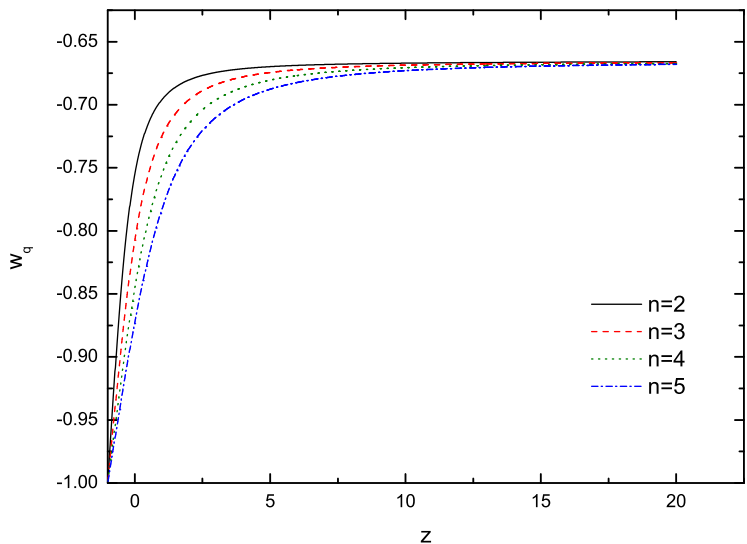}
 \caption{\label{fig1}  The evolution of the equation of state of
 the new agegraphic dark energy with different $n$. Here we use the initial
 condition $\Omega_q(z_{ini})=n^2(1+z_{ini})^{-2}/4$ at
 $z_{ini}=2000$.}
 \end{figure}
 \end{center}

It is worth noting that the new agegraphic dark energy model is a
single-parameter model. As mentioned in \cite{Wei:2007xu}, if $n$ is
given, we can get $\Omega_q(z)$ from Eq.~(\ref{eq10}) with the
initial condition $\Omega_q(z_{ini})=n^2(1+z_{ini})^{-2}/4$. The
initial condition can be chosen at, for example, $z_{ini}=2000$, as
in~\cite{Wei:2007xu}. After that, other physical quantities, such as
$\Omega_m(z)=1-\Omega_q(z)$ and $w_q(z)$, can be obtained. We plot
the evolution curves of $w_{q}$ in Fig.~\ref{fig1} with different
parameter $n$, choosing redshift range $-1< z\leq20$. From this
figure, we can see that at later times ($z\rightarrow -1$ and
$\Omega_q\rightarrow1$), we have $w_{q}\rightarrow -1$, the new
agegraphic dark energy mimics a cosmological constant. When $z=20$,
we have $w_{q}\rightarrow -2/3$, no matter which value of $n$ we
choose.

Next, we shall reconstruct the tachyon scalar-field from the new
agegraphic dark energy. The rolling tachyon condensate in a class of
string theories may have interesting cosmological consequences. It
has been shown by Sen \cite{tachyon} that the decay of D-branes
produces a pressureless gas with finite energy density that
resembles classical dust. A rolling tachyon has an interesting
equation of state whose parameter smoothly interpolates between $-1$
and $0$ \cite{tachyon2}. Thus, tachyon can be viewed as a suitable
candidate for the inflaton at high energy \cite{tachinfl}.
Meanwhile, the tachyon can also act as a source of dark energy
depending upon the form of the tachyon potential \cite{tachde}. The
effective Lagrangian for the tachyon on a non-BPS D3-brane is
described by
\begin{equation}
S=-\int d^4xV(\phi)\sqrt{-{\rm
det}(g_{ab}+\partial_a\phi\partial_b\phi)},
\end{equation}
where $V(\phi)$ is the tachyon potential. The corresponding energy
momentum tensor has the form
\begin{equation}
T_{\mu\nu}={V(\phi)\partial_\mu\phi\partial_\nu\phi\over
\sqrt{1+g^{\alpha\beta}\partial_\alpha\phi\partial_\beta\phi}}
-g_{\mu\nu}V(\phi)\sqrt{1+g^{\alpha\beta}\partial_\alpha\phi\partial_\beta\phi}.
\end{equation} In a flat FRW background the energy density and pressure for the
tachyon field are as following
\begin{equation}
\rho_{\phi}=-T^0_0={V(\phi)\over \sqrt{1-\dot{\phi}^2}},\label{eq12}
\end{equation}
\begin{equation}
p_{\phi}=T^i_i=-V(\phi)\sqrt{1-\dot{\phi}^2}.\label{eq13}
\end{equation}
Consequently the equation of state of the tachyon is given by
\begin{equation}
w_{\phi}=p_{\phi}/\rho_{\phi}=\dot{\phi}^2-1.\label{eq14}
\end{equation}
We see that irrespective the steepness of the tachyon potential, the
equation of state varies between 0 and $-1$. Clearly, the tachyonic
scalar field cannot realize the equation of state crossing $-1$.

Imposing the feature of NADE to the tachyon scalar-field, we should
identify $\rho_\phi$ with $\rho_q$. In following we do not
distinguish the subscripts `$\phi$' and `$q$'. From Eq.~(\ref{eq7}),
we obtain
\begin{equation}
H(z)=H_0\left(\Omega_{m0}(1+z)^3\over 1-\Omega_{
q}\right)^{1/2}.\label{eq15}
\end{equation}
Using Eqs. (\ref{eq12}), (\ref{eq13}) and (\ref{eq15}), the tachyon
potential can be written as
\begin{equation}
{V(\phi)\over \rho_{c0}}={\Omega_{q}\Omega_{m0} (1+z)^3\over
1-\Omega_{q}}\sqrt{-w_{q}},\label{eq16}
\end{equation}
where $\rho_{c0}=3M_p^2H_0^2$ is today's critical density of the
universe. Moreover, using Eqs. (\ref{eq14}) and (\ref{eq15}), the
derivative of the tachyon scalar-field $\phi$ with respect to the
redshift $z$ can be given
\begin{equation}
{\phi'\over H_0^{-1}}=\pm\sqrt{{(1-\Omega_{q})(1+w_{q})\over
\Omega_{m0}(1+z)^5}},\label{eq17}
\end{equation}
where the plus/minus sign is actually arbitrary since it can be
changed by a redefinition of the field, $\phi\rightarrow -\phi$.
Consequently, we can easily obtain the evolutionary form of the
tachyon field
\begin{equation}
\phi(z)=\int\limits_0^z\phi'dz,\label{eq18}
\end{equation}
by fixing the field amplitude at the present epoch ($z=0$) to be
zero, $\phi(0)=0$.
 \begin{center}
 \begin{figure}[htbp]
 \centering
 \includegraphics[width=0.6\textwidth]{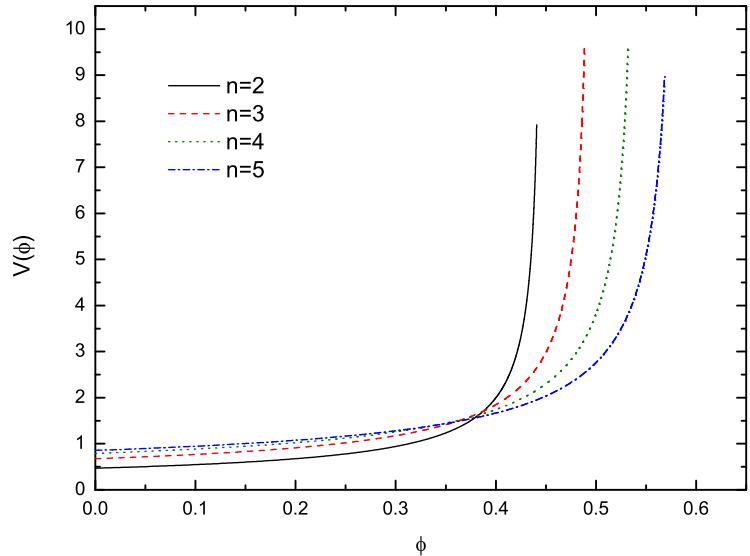}
 \caption{\label{fig2}  The reconstructed tachyon potential from the new agegraphic
 dark energy model, where $\phi$ is in unit of $H_0^{-1}$ and $V(\phi)$ in $\rho_{co}$.}
 \end{figure}
 \end{center}
 \begin{center}
 \begin{figure}[htbp]
 \centering
 \includegraphics[width=0.49\textwidth]{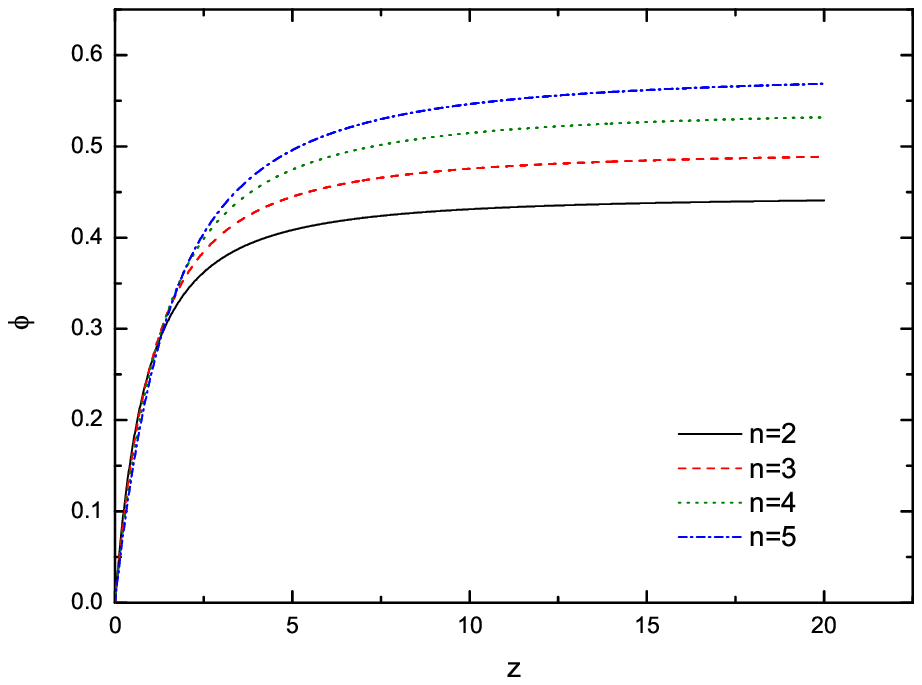}\hfill
 \includegraphics[width=0.49\textwidth]{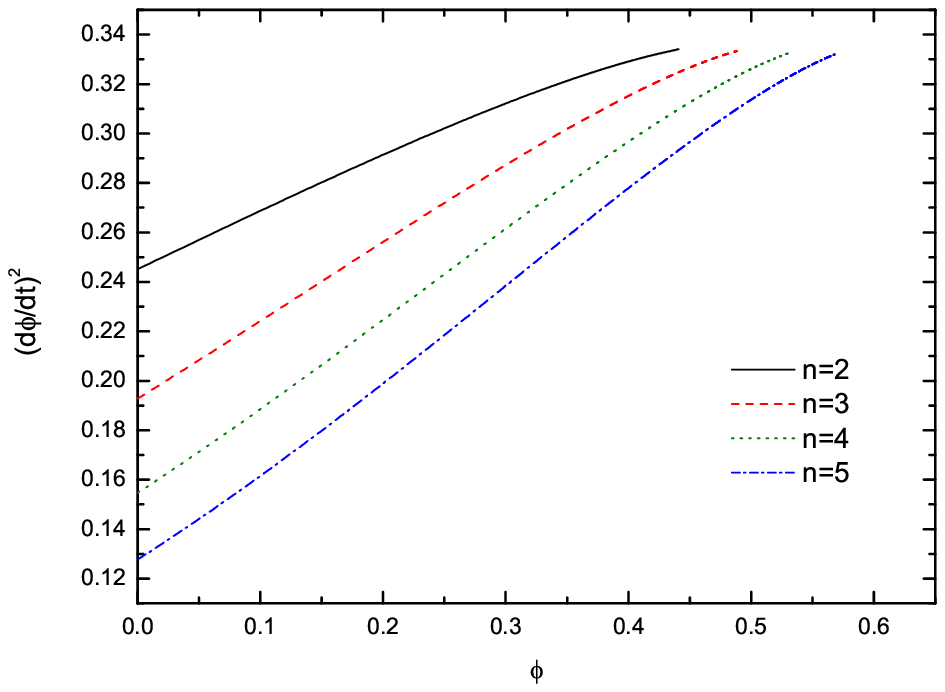}
 \caption{\label{fig3}  The dynamics of the tachyon scalar-field,
 reconstructed according to the evolution of the new agegraphic dark energy.
 Left panel: the evolution of the tachyon field $\phi(z)$, where $\phi$ is in unit of
$H_0^{-1}$; right panel: the kinetic energy density as a function of
scalar-field, namely $\dot{\phi}^2(\phi)$. }
 \end{figure}
 \end{center}

The tachyon models with different potential forms have been
discussed widely in the literature. For the tachyon model
constructed from the NADE, the potential, $V(\phi)$, as well as the
dynamical evolution of the field, $\phi(z)$, can be determined by
Eqs.~(\ref{eq16})$-$(\ref{eq18}). The analytical form of the
potential $V(\phi)$ cannot be derived due to the complexity of these
equations, but we can obtain the tachyon potential numerically.
Using the numerical method, the tachyon potential $V(\phi)$ is
plotted in Fig.~\ref{fig2}, where $\phi(z)$ is also obtained
according to Eqs.~(\ref{eq17}) and (\ref{eq18}), also displayed in
the left panel of Fig.~\ref{fig3}. In the right panel of
Fig.~\ref{fig3}, we plot the kinetic energy density as a function of
the field, namely $\dot{\phi}^2(\phi)$. Actually, Fig.~\ref{fig3}
shows the dynamics of the tachyon scalar-field, reconstructed
according to the evolution of the new agegraphic dark energy.
Selected curves are plotted for the cases of $n= 2,~3,~4$ and $5$.
From Fig.~\ref{fig2}, it is clear to see that the reconstructed
tachyon potential is steeper in the early epoch ($z > 2.5 $) and
becomes very flat near today, according to the evolution of NADE.
Consequently, the tachyon scalar-field $\phi$ rolls down the
potential with the kinetic energy $\dot{\phi}^2$ gradually
decreasing, as indicated in Fig.~\ref{fig3}. The equation of state
of the tachyon $w_{\phi}$, accordingly, decreases gradually with the
cosmic evolution, and as a result, $dw_{\phi}/d\ln a<0$.

 \begin{center}
 \begin{figure}[tbhp]
 \centering
 \includegraphics[width=0.49\textwidth]{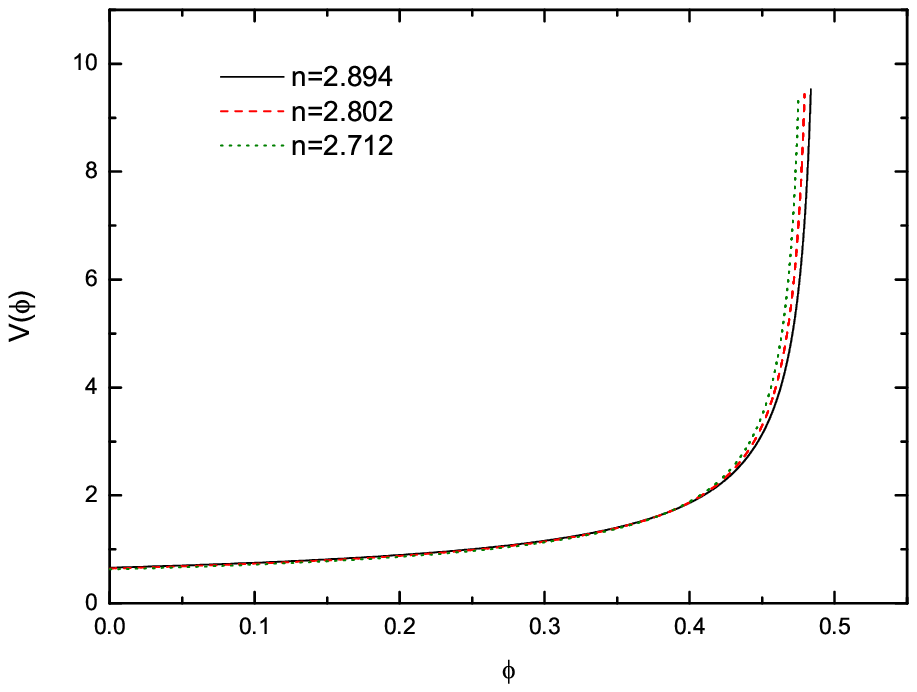}\hfill
 \includegraphics[width=0.49\textwidth]{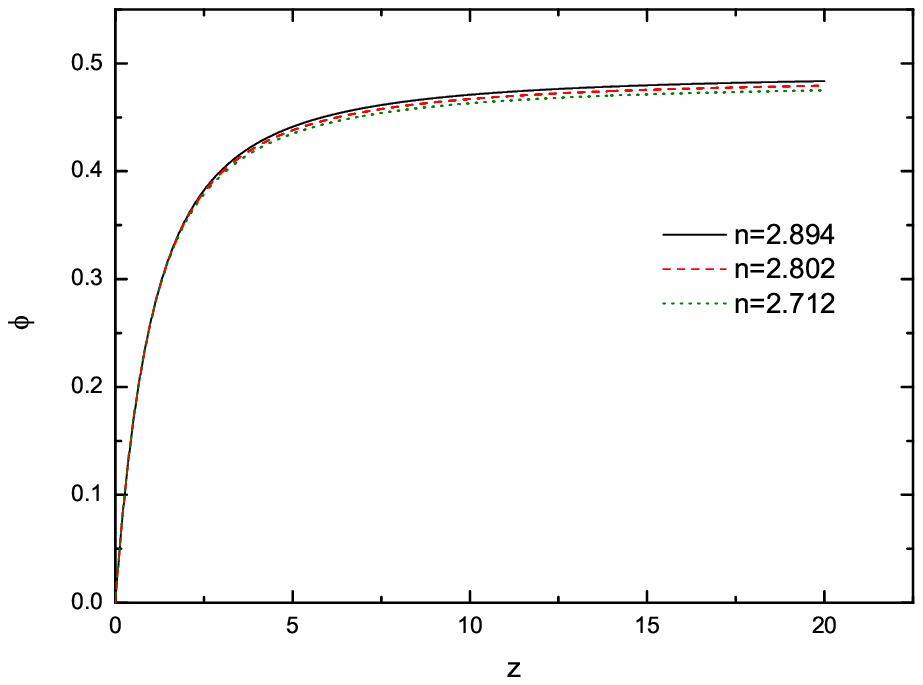}
 \caption{\label{fig4}
 Reconstruction of the tachyon model, according to the evolution of the new
 agegraphic dark energy. Here we take $n=2.802^{+0.092}_{-0.090}$, from the
 cosmological constraints on the new agegraphic dark energy.}
 \end{figure}
 \end{center}

Recently Wei~\cite{Wei:2008rv} considered the cosmological
constraints on the new agegraphic dark energy model by using the
latest observational data. When combining the 307 Union type Ia
supernovae (SNIa), the 32 calibrated Gamma-Ray Bursts (GRBs) at
$z>1.4$, the updated shift parameter $R$ from WMAP 5-year data
(WMAP5), and the distance parameter $A$ of the measurement of baryon
acoustic oscillation (BAO) peak in the distribution of SDSS luminous
red galaxies with the updated scalar spectral index $n_s$ from
WMAP5, the fitting for the new agegraphic dark energy model gives
the parameter constraint in $1\sigma$: $n=2.802^{+0.092}_{-0.090}$,
with $\chi^2_{min}=336.061$. According to
Eqs.~(\ref{eq16})$-$(\ref{eq18}), the reconstructed tachyon
potential $V(\phi)$ and field $\phi(z)$ are plotted in
Fig.~\ref{fig4}, with the cases of $n=2.802^{+0.092}_{-0.090}$.

In summary, we suggest a correspondence between the new agegraphic
dark energy scenario and the tachyon model in this paper. The new
agegraphic dark energy model was proposed by Wei and
Cai~\cite{Cai2}, based on the K\'{a}rolyh\'{a}zy uncertainty
relation, which arises from the quantum mechanics together with
general relativity. We adopt the viewpoint that the scalar field
models of dark energy are effective theories of an underlying theory
of dark energy. If we regard the scalar-field model as an effective
description of such a theory, we should be capable of using the
scalar-field model to mimic the evolving behavior of the new
agegraphic dark energy and reconstructing this scalar-field model
according to the evolutionary behavior of  new agegraphic dark
energy. We show that the new agegraphic dark energy can be described
totally by the tachyon in a certain way. A correspondence between
the new agegraphic dark energy and tachyon field has been
established, and the tachyon potential as well as the dynamics of
the tachyon field have been reconstructed.

\section*{Acknowledgements}

This work was supported by the National Natural Science Foundation
of China (Grant No. 10705041).



\begin{thebibliography}{}

\bibitem{Riess98}
A.~G.~Riess {\it et al.}  [Supernova Search Team Collaboration],
  Astron.\ J.\  {\bf 116} (1998) 1009
  [arXiv:astro-ph/9805201];
  S.~Perlmutter {\it et al.}  [Supernova Cosmology Project Collaboration],
  Astrophys.\ J.\  {\bf 517}, 565 (1999)
  [astro-ph/9812133].

\bibitem{Tegmark04}
M.~Tegmark {\it et al.}  [SDSS Collaboration],
  Phys.\ Rev.\  D {\bf 69} (2004) 103501
  [arXiv:astro-ph/0310723].

\bibitem{Spergel03}
D.~N.~Spergel {\it et al.}  [WMAP Collaboration],
  Astrophys.\ J.\ Suppl.\  {\bf 148} (2003) 175
  [arXiv:astro-ph/0302209].

\bibitem{lamada}
T.~Padmanabhan, Phys.\ Rept.\  {\bf 380}, 235 (2003)
[hep-th/0212290]; S.~M.~Carroll, astro-ph/0310342; R.~Bean,
S.~Carroll and M.~Trodden, astro-ph/0510059; V.~Sahni and
A.~A.~Starobinsky,
 Int.\ J.\ Mod.\ Phys.\ D {\bf 9}, 373 (2000) [astro-ph/9904398];
T.~Padmanabhan, Curr.\ Sci.\  {\bf 88}, 1057 (2005)
[astro-ph/0411044]; S.~Nobbenhuis,
 Found.\ Phys.\  {\bf 36}, 613 (2006) [gr-qc/0411093];
E.~J.~Copeland, M.~Sami and S.~Tsujikawa,
 Int.\ J.\ Mod.\ Phys.\  D {\bf 15}, 1753 (2006)
 [hep-th/0603057];
A.~Albrecht {\it et al.}, astro-ph/0609591; R.~Trotta and R.~Bower,
astro-ph/0607066; M.~Kamionkowski, arXiv:0706.2986 [astro-ph];
B.~Ratra and M.~S.~Vogeley, arXiv:0706.1565 [astro-ph];
E.~V.~Linder, arXiv:0705.4102 [astro-ph]; P.~J.~Steinhardt, in {\it
Critical Problems in Physics}, edited by V.~L.~Fitch and
D.~R.~Marlow (Princeton University Press, Princeton, NJ, 1997).

\bibitem{quintessence}
  P.~J.~E.~Peebles and B.~Ratra,
  Astrophys.\ J.\  {\bf 325} L17 (1988);
  B.~Ratra and P.~J.~E.~Peebles,
  Phys.\ Rev.\ D {\bf 37} 3406 (1988);
  C.~Wetterich,
  Nucl.\ Phys.\ B {\bf 302} 668 (1988);
  J.~A.~Frieman, C.~T.~Hill, A.~Stebbins and I.~Waga,
  Phys.\ Rev.\ Lett.\  {\bf 75}, 2077 (1995)
  [astro-ph/9505060];
  M.~S.~Turner and M.~J.~White,
  Phys.\ Rev.\ D {\bf 56}, 4439 (1997)
  [astro-ph/9701138];
  A.~R.~Liddle and R.~J.~Scherrer,
  Phys.\ Rev.\ D {\bf 59}, 023509 (1999)
  [astro-ph/9809272];
  I.~Zlatev, L.~M.~Wang and P.~J.~Steinhardt,
  Phys.\ Rev.\ Lett.\  {\bf 82}, 896 (1999)
  [astro-ph/9807002];
  P.~J.~Steinhardt, L.~M.~Wang and I.~Zlatev,
  Phys.\ Rev.\ D {\bf 59}, 123504 (1999)
  [astro-ph/9812313].


\bibitem{kessence}
  C.~Armendariz-Picon, V.~F.~Mukhanov and P.~J.~Steinhardt,
  Phys.\ Rev.\ Lett.\  {\bf 85}, 4438 (2000)
  [astro-ph/0004134];
  C.~Armendariz-Picon, V.~F.~Mukhanov and P.~J.~Steinhardt,
  Phys.\ Rev.\ D {\bf 63}, 103510 (2001)
  [astro-ph/0006373].



\bibitem{phantom}
  R.~R.~Caldwell,
  Phys.\ Lett.\ B {\bf 545}, 23 (2002)
  [astro-ph/9908168].


\bibitem{tachyon}
  A.~Sen,
  JHEP {\bf 0207}, 065 (2002)
  [hep-th/0203265].


\bibitem{quintom}
  B.~Feng, X.~L.~Wang and X.~M.~Zhang,
  Phys.\ Lett.\ B {\bf 607}, 35 (2005)
  [astro-ph/0404224];
  Z.~K.~Guo, Y.~S.~Piao, X.~M.~Zhang and Y.~Z.~Zhang,
  Phys.\ Lett.\ B {\bf 608}, 177 (2005)
  [astro-ph/0410654];
  X.~Zhang,
  Commun.\ Theor.\ Phys.\  {\bf 44}, 762 (2005).

\bibitem{ghost}
  N.~Arkani-Hamed, H.~C.~Cheng, M.~A.~Luty and S.~Mukohyama,
  JHEP {\bf 0405}, 074 (2004)
  [hep-th/0312099];
  S.~Mukohyama,
  JCAP {\bf 0610}, 011 (2006)
  [hep-th/0607181];
  F.~Piazza and S.~Tsujikawa,
  JCAP {\bf 0407}, 004 (2004)
  [hep-th/0405054];
  J.~Zhang, X.~Zhang and H.~Liu,
  Mod.\ Phys.\ Lett.\  A {\bf 23}, 139 (2008)
  [arXiv:astro-ph/0612642].

\bibitem{intde}
  L.~Amendola,
  Phys.\ Rev.\ D {\bf 62}, 043511 (2000)
  [astro-ph/9908023];
  D.~Comelli, M.~Pietroni and A.~Riotto,
  Phys.\ Lett.\ B {\bf 571}, 115 (2003)
  [hep-ph/0302080];
  X.~Zhang,
  Mod.\ Phys.\ Lett.\ A {\bf 20}, 2575 (2005)
  [astro-ph/0503072];
  X.~Zhang,
  Phys.\ Lett.\  B {\bf 611}, 1 (2005)
  [astro-ph/0503075].

\bibitem{brane}
  C.~Deffayet, G.~R.~Dvali and G.~Gabadadze,
  Phys.\ Rev.\ D {\bf 65}, 044023 (2002)
  [astro-ph/0105068];
  V.~Sahni and Y.~Shtanov,
  JCAP {\bf 0311}, 014 (2003)
  [astro-ph/0202346].

\bibitem{cg}
  A.~Y.~Kamenshchik, U.~Moschella and V.~Pasquier,
  Phys.\ Lett.\ B {\bf 511}, 265 (2001)
  [gr-qc/0103004];
  M.~C.~Bento, O.~Bertolami and A.~A.~Sen,
  Phys.\ Rev.\  D {\bf 66}, 043507 (2002)
  [gr-qc/0202064];
  X.~Zhang, F.~Q.~Wu and J.~Zhang,
  JCAP {\bf 0601}, 003 (2006)
  [astro-ph/0411221].

\bibitem{Li04}
M.~Li,
  Phys.\ Lett.\  B {\bf 603} (2004) 1
  [arXiv:hep-th/0403127].

\bibitem{holofit}
  X.~Zhang and F.~Q.~Wu,
  Phys.\ Rev.\  D {\bf 72}, 043524 (2005)
  [arXiv:astro-ph/0506310];
  X.~Zhang and F.~Q.~Wu,
  Phys.\ Rev.\  D {\bf 76}, 023502 (2007)
  [arXiv:astro-ph/0701405];
  Q.~G.~Huang and Y.~G.~Gong,
  JCAP {\bf 0408}, 006 (2004)
  [arXiv:astro-ph/0403590];
  Z.~Chang, F.~Q.~Wu and X.~Zhang,
  Phys.\ Lett.\  B {\bf 633}, 14 (2006)
  [arXiv:astro-ph/0509531];
  J.~Y.~Shen, B.~Wang, E.~Abdalla and R.~K.~Su,
  Phys.\ Lett.\  B {\bf 609}, 200 (2005)
  [arXiv:hep-th/0412227];
  Z.~L.~Yi and T.~J.~Zhang,
  Mod.\ Phys.\ Lett.\  A {\bf 22}, 41 (2007)
  [arXiv:astro-ph/0605596];
  Y.~Z.~Ma, Y.~Gong and X. Chen,
  arXiv:0711.1641 [astro-ph];
  Q.~Wu, Y.~Gong, A.~Wang and J.~S.~Alcaniz,
  Phys.\ Lett.\  B {\bf 659}, 34 (2008)
  [arXiv:0705.1006 [astro-ph]];
  X.~Zhang,
  Int.\ J.\ Mod.\ Phys.\  D {\bf 14}, 1597 (2005)
  [arXiv:astro-ph/0504586];
  M.~R.~Setare, J.~Zhang and X.~Zhang,
  JCAP {\bf 0703}, 007 (2007)
  [arXiv:gr-qc/0611084];
  J.~Zhang, X.~Zhang and H.~Liu,
  Phys.\ Lett.\  B {\bf 659}, 26 (2008)
  [arXiv:0705.4145 [astro-ph]];
  J.~Zhang, X.~Zhang and H.~Liu,
  Eur.\ Phys.\ J.\  C {\bf 52}, 693 (2007)
  [arXiv:0708.3121 [hep-th]].






\bibitem{Cai1}
R. G. Cai, Phys. Lett. B $\textbf{657}$, 228-231 (2007)
[arXiv:hep-th/0707.4049].

\bibitem{Cai2}
H. Wei, R. G. Cai, Phys. Lett. B $\textbf{660}$ 113 (2008)
[arXiv:astro-ph/0708.0884].

\bibitem{holorec}
  X.~Zhang,
  Phys.\ Lett.\  B {\bf 648}, 1 (2007)
  [arXiv:astro-ph/0604484];
  X.~Zhang,
  Phys.\ Rev.\  D {\bf 74}, 103505 (2006)
  [arXiv:astro-ph/0609699];
  J.~Zhang, X.~Zhang and H.~Liu,
  Phys.\ Lett.\  B {\bf 651}, 84 (2007)
  [arXiv:0706.1185 [astro-ph]];
  Y.~Z.~Ma and X.~Zhang,
  Phys.\ Lett.\  B {\bf 661}, 239 (2008)
  [arXiv:0709.1517 [astro-ph]];
  N.~Cruz, P.~F.~Gonzalez-Diaz, A.~Rozas-Fernandez and G.~Sanchez,
  arXiv:0812.4856 [gr-qc];
  I.~P.~Neupane,
  Phys.\ Rev.\  D {\bf 76}, 123006 (2007)
  [arXiv:0709.3096 [hep-th]];
  J.~Zhang, X.~Zhang and H.~Liu,
  Eur.\ Phys.\ J.\  C {\bf 54}, 303 (2008)
  [arXiv:0801.2809 [astro-ph]];
  J.~P.~Wu, D.~Z.~Ma and Y.~Ling,
  Phys.\ Lett.\  B {\bf 663}, 152 (2008)
  [arXiv:0805.0546 [hep-th]];
  X.~Zhang,
  arXiv:0901.2262 [astro-ph.CO];
  C.~J.~Feng,
  arXiv:0810.2594 [hep-th];
  X.~Wu and Z.~H.~Zhu,
  Phys.\ Lett.\  B {\bf 660}, 293 (2008)
  [arXiv:0712.3603 [astro-ph]];
  M.~R.~Setare,
  Phys.\ Lett.\  B {\bf 648}, 329 (2007)
  [arXiv:0704.3679 [hep-th]].






\bibitem{r1}
F.~K\'{a}rolyh\'{a}zy, Nuovo Cim.\  A {\bf 42}, 390 (1966);
F.~K\'{a}rolyh\'{a}zy, A.~Frenkel and B.~Luk\'{a}cs,
 in {\it Physics as Natural Philosophy}, edited by
 A.~Simony and H.~Feschbach, MIT Press, Cambridge, MA (1982);
F.~K\'{a}rolyh\'{a}zy, A.~Frenkel and B.~Luk\'{a}cs,
 in {\it Quantum Concepts in Space and Time}, edited by
 R.~Penrose and C.~J.~Isham, Clarendon Press, Oxford (1986).

\bibitem{r2}
M.~Maziashvili,
  Int.\ J.\ Mod.\ Phys.\  D {\bf 16}, 1531 (2007)
  [arXiv:gr-qc/0612110].

\bibitem{r3}
M.~Maziashvili,
  Phys.\ Lett.\  B {\bf 652}, 165 (2007)
  [arXiv:0705.0924 [gr-qc]].


\bibitem{Wei:2007xu}
  H.~Wei and R.~G.~Cai,
  Phys.\ Lett.\  B {\bf 663}, 1 (2008)
  [arXiv:0708.1894 [astro-ph]].

\bibitem{tachyon2}
  G.~W.~Gibbons,
  Phys.\ Lett.\  B {\bf 537}, 1 (2002)
  [hep-th/0204008].

\bibitem{tachinfl}
  A.~Mazumdar, S.~Panda and A.~Perez-Lorenzana,
  Nucl.\ Phys.\  B {\bf 614}, 101 (2001)
  [hep-ph/0107058];
  A.~Feinstein,
  Phys.\ Rev.\  D {\bf 66}, 063511 (2002)
  [hep-th/0204140];
  Y.~S.~Piao, R.~G.~Cai, X.~M.~Zhang and Y.~Z.~Zhang,
  Phys.\ Rev.\  D {\bf 66}, 121301 (2002)
  [hep-ph/0207143].

\bibitem{tachde}
  T.~Padmanabhan,
  Phys.\ Rev.\  D {\bf 66}, 021301 (2002)
  [hep-th/0204150];
  J.~S.~Bagla, H.~K.~Jassal and T.~Padmanabhan,
  Phys.\ Rev.\  D {\bf 67}, 063504 (2003)
  [astro-ph/0212198];
  L.~R.~W.~Abramo and F.~Finelli,
  Phys.\ Lett.\  B {\bf 575}, 165 (2003)
  [astro-ph/0307208];
  J.~M.~Aguirregabiria and R.~Lazkoz,
  Phys.\ Rev.\  D {\bf 69}, 123502 (2004)
  [hep-th/0402190];
  Z.~K.~Guo and Y.~Z.~Zhang,
  JCAP {\bf 0408}, 010 (2004)
  [hep-th/0403151];
  E.~J.~Copeland, M.~R.~Garousi, M.~Sami and S.~Tsujikawa,
  Phys.\ Rev.\  D {\bf 71}, 043003 (2005)
  [hep-th/0411192].




\bibitem{Wei:2008rv}
  H.~Wei,
  arXiv:0809.0057 [astro-ph].




















\end{thebibliography}
\end{document}